# Designing an ultra negative dispersion Photonic Crystal Fiber (PCFs) with square lattice geometry

Partha Sona Maji* and Partha Roy Chaudhuri

Department of Physics & Meteorology, Indian Institute of Technology Kharagpur-721 302, INDIA

*Tel: +91-3222-283842 Fax: +91-3222-255303,*

**Corresponding author: parthamaji@phy.iitkgp.ernet.in*

**Abstract:** In this article we have theoretically investigated the dispersion characteristics of dual-core PCF, based on square-lattice geometry by varying different parameters. The fiber exhibits a very large negative dispersion because of rapid slope change of the refractive indices at the coupling wavelength between the inner core and outer core. The dependence of different geometrical parameters namely hole-to-hole spacing (Λ) and different air-hole diameter (*d*) was investigated in detail. By proper adjustment of the available parameters, a high negative dispersion value of -47,500 *ps/nm/km* has been achieved around the wavelength of 1550nm. Our proposed fiber will be an excellent device for dispersion compensation in long-haul data transmission as being thousand times more than the available DCFs.

## 1. INTRODUCTION

Photonic crystal fibers (PCFs)[**1,2**] or Holey Optical Fibers offered a tremendous variety of possible geometries utilizing the *shape*, *size* and *positioning* of air-holes in the micro-structured cladding. The air-hole diameter (*d*) and hole-to-hole spacing (Λ) not only control the dispersion properties, but also the transmission and the nonlinear properties of the fiber as well. Achieving very high negative values of dispersion around the communication band has been the target for a long time [**3-15**].

The principle behind having a very large negative dispersion in these Dispersion Compensating Fibers (DCFs) being the coupling between two spatially separated asymmetric concentric cores which support two leaky modes: inner mode and outer mode. By proper design, mode matching can take place between these two modes at the desired wavelength. A few analyses have been performed to realize high negative dispersion with triangular lattice PCF [**8-15**]. In this work we have studied rigorously towards achieving high negative dispersion value with regular square lattice. Square lattice based PCF is superior to triangular-lattice PCF for certain properties [**16-17**]. Square lattice PCF shows wider range of single mode operation with the same *d*/Λ value compared to the triangular one [**16**]. The effective area of square-lattice PCF is higher than triangular one, making the former better for high power management [**17**]. Square-lattice PCF can better compensate the in-line dispersion around the 1550nm wavelength, than the triangular-lattice PCF [**17**].

In recent times, a square-lattice PCF preform has been realized with a standard fabrication process, stack and draw, in order to study the localization and control of high frequency sound by introducing two solid defects in the periodic distribution of air-holes [**18**]. Thus the technological feasibility of the square-lattice PCFs has been demonstrated, since the final PCFs can be obtained by drawing the intermediate prepared preforms [**18**]. In another example, experimental study of negative refraction has been studied with square-lattice Photonic Crystal [**19**]. So, square-lattice PCF can be experimentally realized like that of the usual triangular-lattice PCF.

## 2. Geometry of the structure and analysis method:

Cross sectional geometry of the proposed/studied fiber has been shows in Fig. 1. It is well known that a triangular lattice PCF is usually described by air-hole diameter *d* and hole-to-hole distance (also called as pitch) Λ. Now, we use Λ as the hole-to-hole spacing both in horizontal and vertical directions in the square-lattice PCF geometry with $d_1$ as the diameter of the bigger air holes. The central air-hole is missing, making it the inner core. The inner cladding is formed by the first two air-hole rings with air-hole diameter $d_1$. The diameter of the air-holes for the 3rd air-hole ring is reduced, thereby increasing the local refractive index of the ring, making the ring as the outer core. The diameter of the air-holes in this outer ring is represented as $d_2$. The rings of holes beyond the third rings form the outer cladding with air-hole diameter $d_1$. The back ground of the fiber is taken to be silica whose refractive index has been considered through Sellmier's Eqn. (1).

$$n^2(\lambda) = A + \frac{B_1\lambda^2}{\lambda^2-\lambda_1^2} + \frac{B_2\lambda^2}{\lambda^2-\lambda_2^2} + \frac{B_3\lambda^2}{\lambda^2-\lambda_3^2} \qquad (1)$$

With $B_1$= 0.696166300, $B_2$=0.407942600 and $B_3$=.897479400 and $\lambda_1$=0.0684043μm, $\lambda_2$=0.1162414μm and $\lambda_3$=9.896161μm [**20**].

We solve the guided modes of the present fiber by the CUDOS MOF Utilities [**21**] that simulate PCFs using the multipole method [**22-23**].We have calculated the dispersion parameter using Eqn. (2)

$$D = -\frac{\lambda}{c}\frac{d^2\,\mathrm{Re}[n_{eff}]}{d\lambda^2} \qquad (2)$$

with $Re(n_{eff})$ is the real part of the effective indices obtained from simulations and c is the speed of light in vacuum.

The confinement loss for the structures has been calculated through Eqn. (3).

$$L = \frac{2\pi}{\lambda}\frac{20}{\ln(10)}10^6\,\mathrm{Im}(n_{eff})\ dB/m \qquad (3)$$

where Im($n_{eff}$) is the imaginary part of the effective indices (obtained from the simulations) and λ in micrometer.

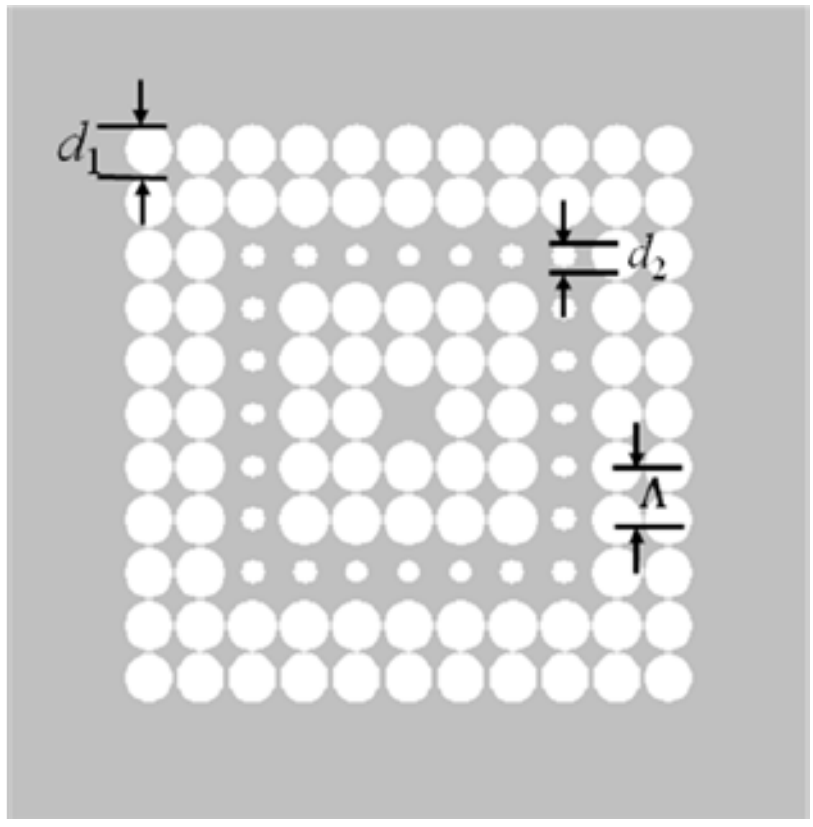

**Fig. 1:** Cross section of the proposed /studied fiber. The air-hole diameter of the third air-hole ring is reduced to create the outer core, thereby creating the dual-core structure.

## 3. Dispersion analysis of the structure:

We started our dispersion analysis with Λ =1.40μm and $d_1$/Λ =0.8 and $d_2/d_1$=0.5. A very high negative dispersion of -21,700 *ps/nm/km* around the wavelength of 1522 nm was observed as shown in Fig. 4. The corresponding variations of effective index have been presented in Fig. 2 which shows a distinctive change of slope at the coupling wavelength. The variation of the indices for both the inner and outer core is presented in the above figure. The cross-off between the two cores (inner core and outer core) can be better viewed from Fig. 3 which represents the imaginary part of the refractive indices (Im($n_{eff}$)) of the two cores. The two curves meet at the coupling wavelength of 1522nm. After the coupling, most of the power in the inner core goes to the outer core. This principle can be used for suppressing spontaneous emission of certain wavelengths.

The dependence of dispersion upon the geometrical parameters ($d_1$, $d_2$ and Λ) has been presented in Figs. 5–7. The dependence of the variation of bigger air-holes ($d_1$) upon dispersion has been presented in Fig. 5. For this purpose we have kept Λ=1.40μm, keeping $d_2/d_1$=0.5. From the figure it is clearly visible that the absolute values of the biggest negative dispersion increase as $d_1$ increases (air filling rate increases), the corresponding coupling wavelength is red-shifted, and the absolute values of the dispersion slope increase but the value of full width at half maximum (FWHM) decreases. From the figure it can be easily observed that with an increase of negative dispersion the corresponding FWHM decreases making the product of bandwidth and peak dispersion almost constant. The dispersion curves of PCFs for different $d_2$ has been presented in Fig. 6 with Λ =1.40μm and $d_1$=1.12μm. The figure clearly represents that values of the biggest negative dispersion decrease as $d_2$ increases (air filling rate of the outer core increases), the corresponding coupling wavelength is red-shifted and the absolute values of the dispersion slope decrease. Here with the increases of outer air-hole diameter the peak dispersion decreases but the corresponding FWHM increases keeping the product of bandwidth and peak dispersion remain almost constant. The dependence of hole-to-hole distance (Λ) upon dispersion has been presented in Fig. 7. For this purpose we have considered two different PCFs with $d_1$/Λ= 0.8 and $d_2/d_1$ = 0.5. Figure 7 clearly represents that the absolute values of the biggest negative dispersion reduces appreciably as Λ increases and the analogous wavelength is red-shifted and the absolute values of the dispersion slope decrease significantly.

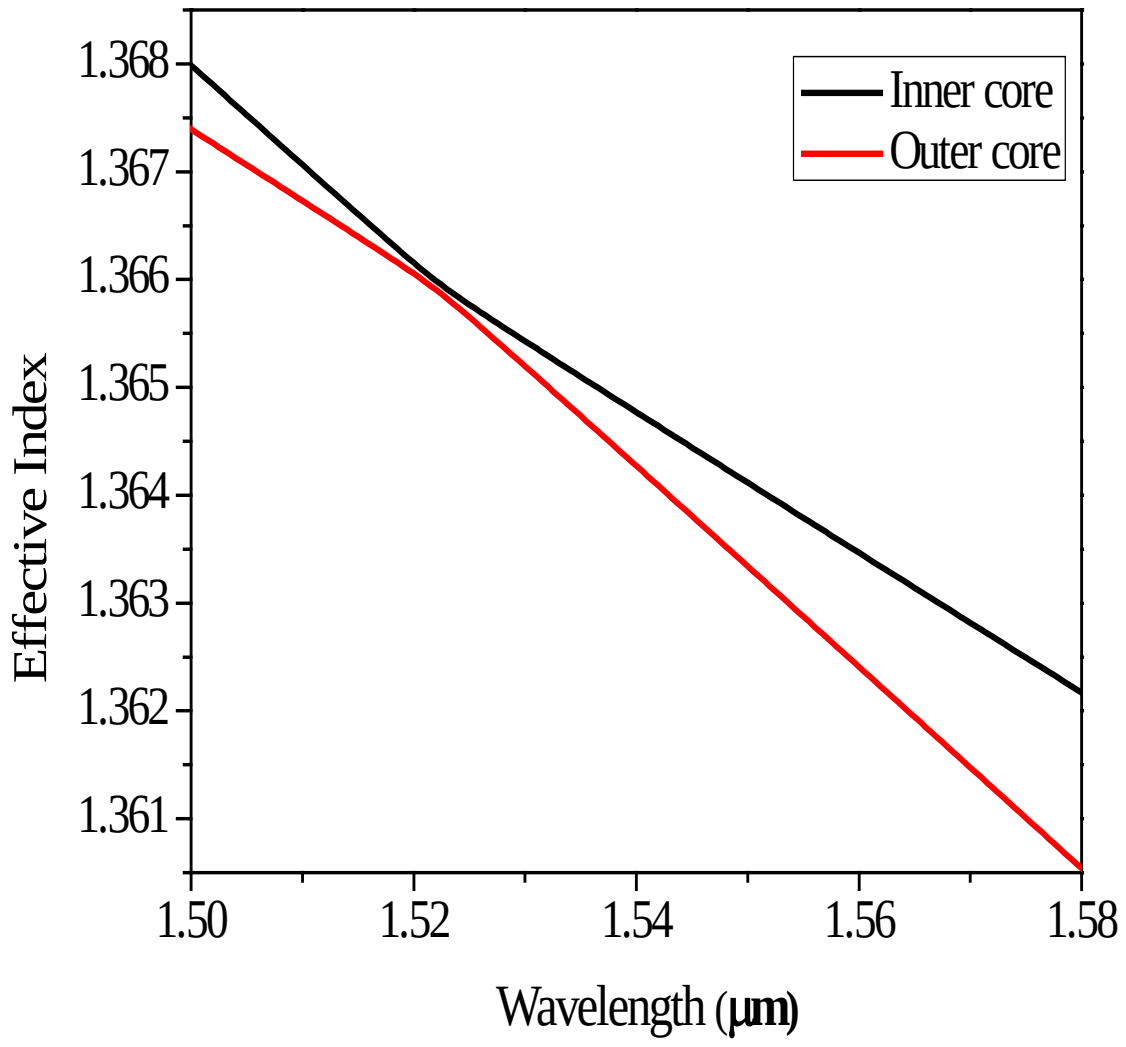

**Fig. 2:** Variation of real part of the effective indices for both the cores (inner core-black line and outer core red-line) of PCFs with Λ=1.40μm, $d_1$=1.12μm and $d_2$=0.56μm.

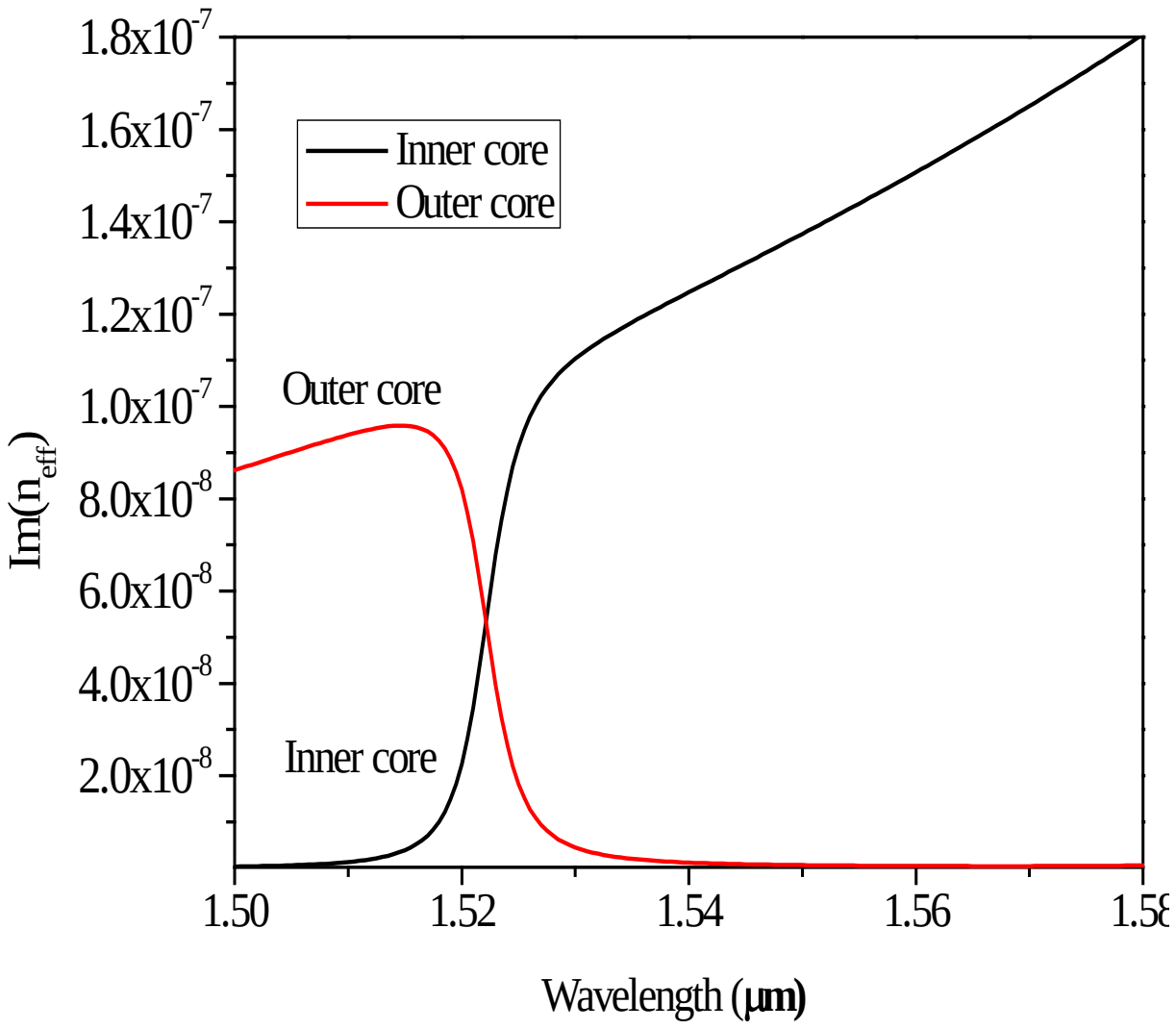

**Fig. 3**: *Im*($n_{eff}$) variation for the two cores of the PCFs with Λ=1.4μm, $d_1$=1.12μm and $d_2$=0.56μm.

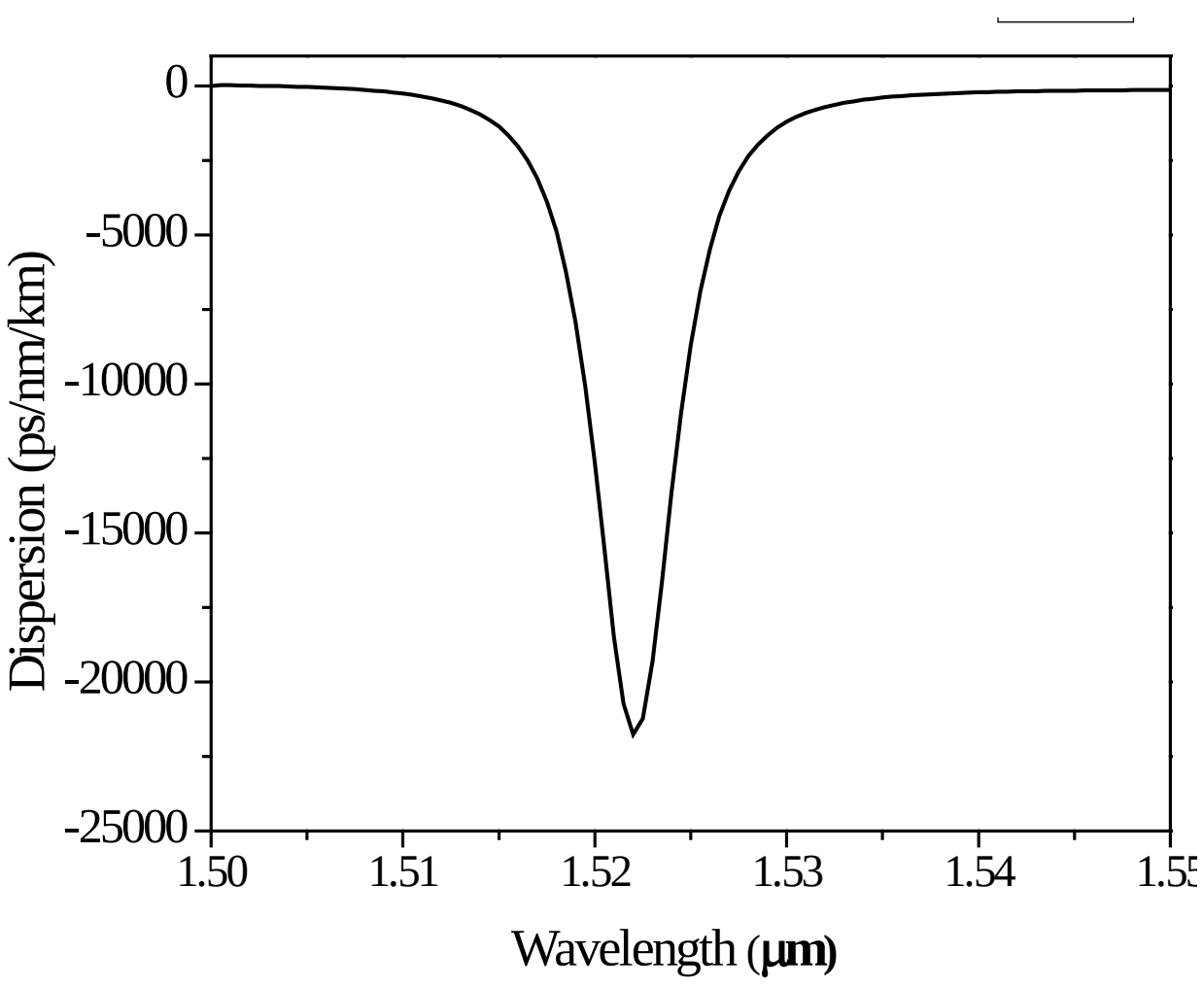

**Fig. 4:** Dispersion curve for PCFs with Λ=1.40μm, $d_1$=1.12μm and $d_2$=0.56μm

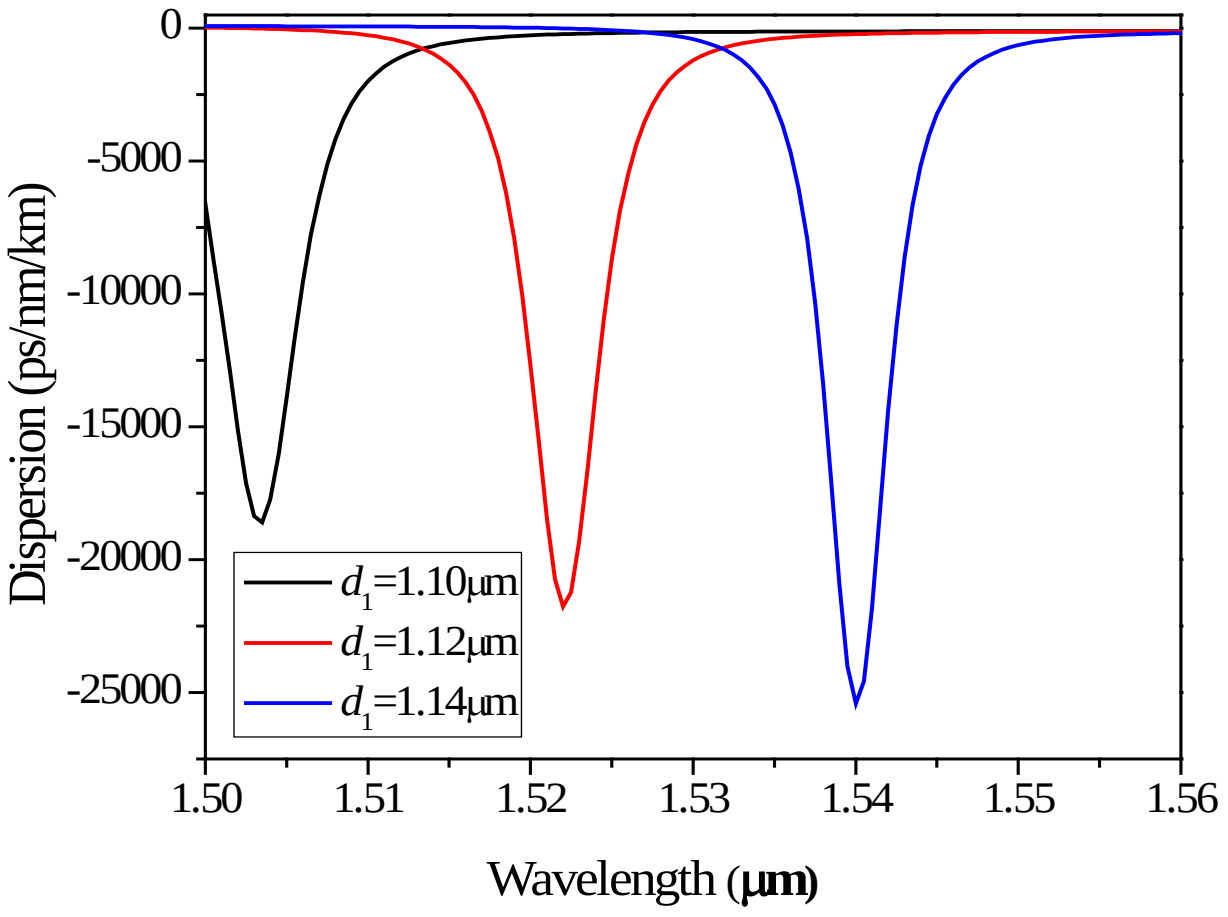

**Fig. 5:** Variation of dispersion for Λ=1.40μm for $d_2/d_1$=0.5 for different values of $d_1$ (1.10μm, 1.12μm and 1.14μm)

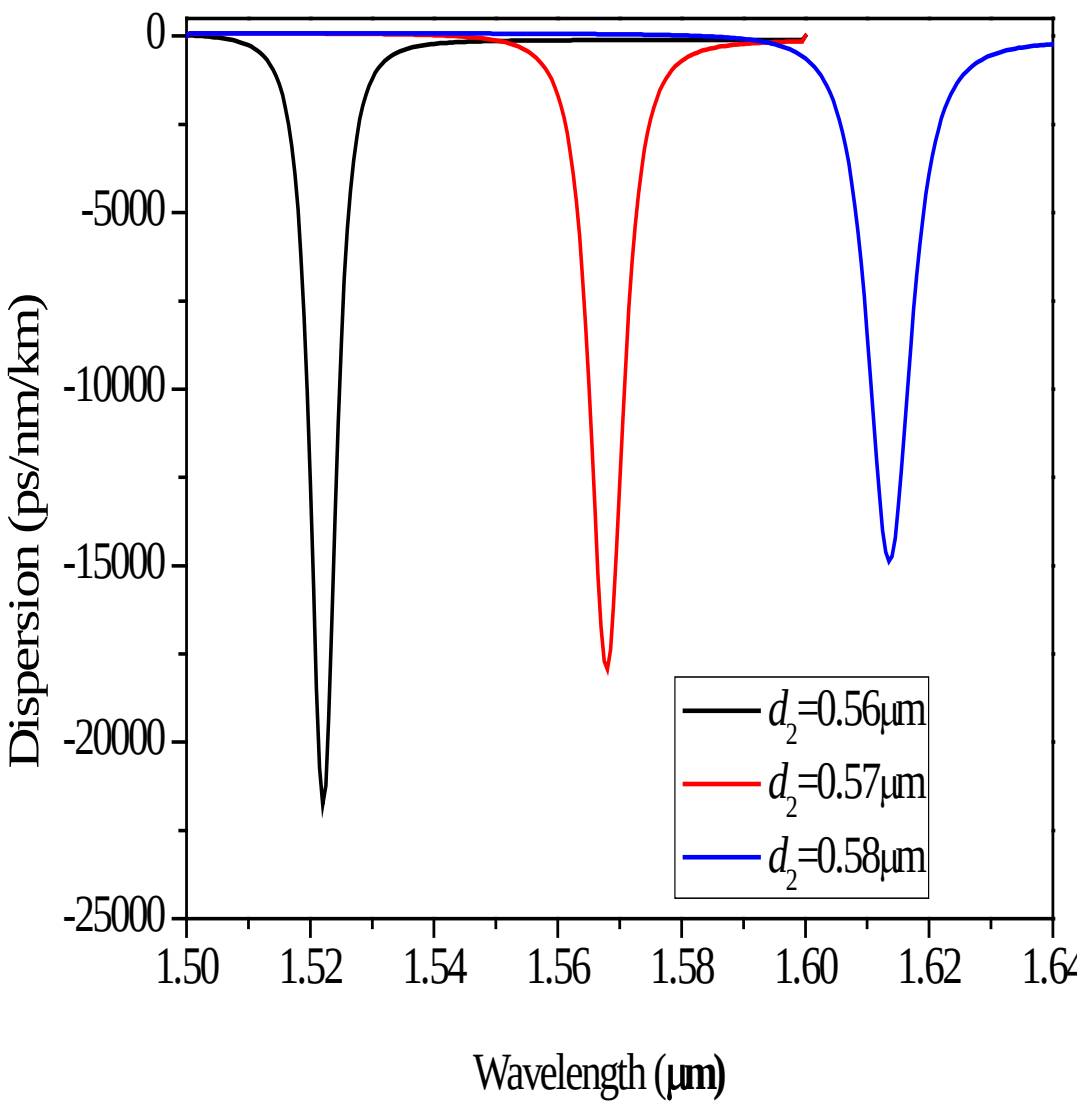

**Fig. 6:** Variation of $D$ for Λ=1.40 μm for different values of $d_2$ (0.55μm, 0.56μm and 0.57μm) with $d_1$=1.12μm.

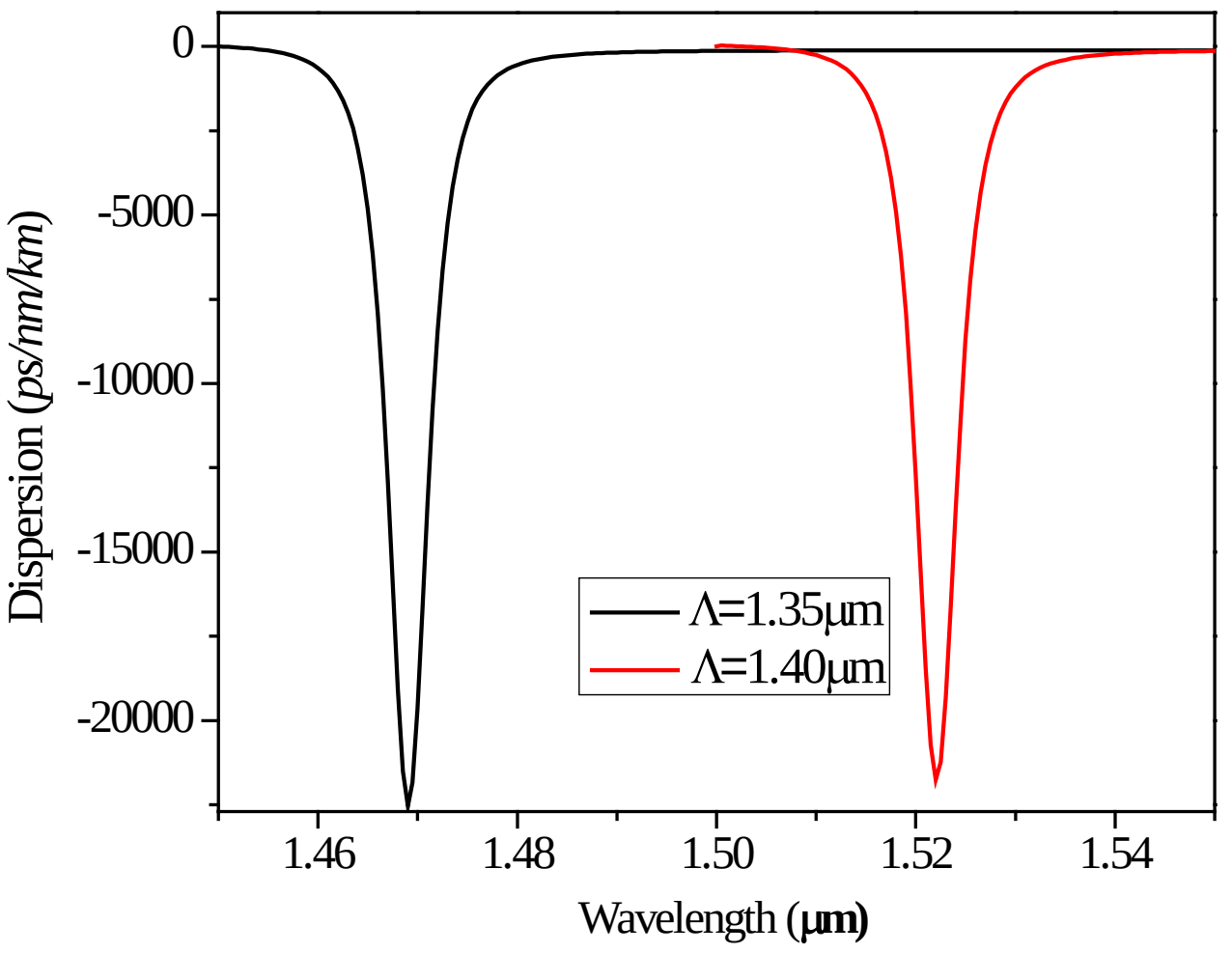

**Fig. 7:** Variation of $D$ for different values of Λ with $d_2/d_1$=0.5 and $d_1$/Λ=0.8.

In this section, we'll study the fabrication tolerance of the dual core S-PCF for designing ultra-low dispersion at the required wavelength. For this purpose, we have considered the S-PCF as demonstrated in Fig. 4 with Λ=1.40μm, $d_1$=1.12μm and $d_2$=0.56μm. The tolerances have been considered for four values including the original values. For this case, we have changed the values of the available parameters (Λ, $d_1$ and $d_2$) with a change from -2% to +2% in a step of 1%. The variation of the dispersion for percentage change of Λ, $d_1$ and $d_2$ have been demonstrated in Fig. 8, Fig. 9 and Fig 10 respectively. The figures reveals that for a decrease of Λ, peak dispersion increases towards a smaller wavelength (Fig. 8(a) ), while peak dispersion increases with an increase of "$d_1$" towards a smaller wavelength (Fig. 9(a)) with dispersion increases with a decrease of "$d_2$" (Fig. 10(a)). The figures (Fig. 8(b), Fig. 9(b) and Fig. 10(b)) reveal that peak dispersion wavelength variation is almost linear with a change of parameters. However, the same can't be concluded for peak dispersion variation (Fig. 8(a), Fig. 9(a) and Fig. 10(a)) with parameters and the variation is close to linear but deviates from linear relationship. One of the important information that can be concluded form the above figures is that the peak dispersion wavelength value is most sensitive to the change of smaller air-hole diameter ($d_2$), while the effect of change of hole-to-hole distance (Λ) is the least. Another interesting fact can be observed that peak dispersion is most sensitive to the change of bigger air-hole diameter ($d_1$), while hole-to-hole distance (Λ) effects the least change. These engineering aspects of effect of change of parameters by a certain percentage towards different peak dispersion and corresponding wavelength will be of great help towards possible fabrication of a dual core SPCF for narrowband ultra negative dispersion.

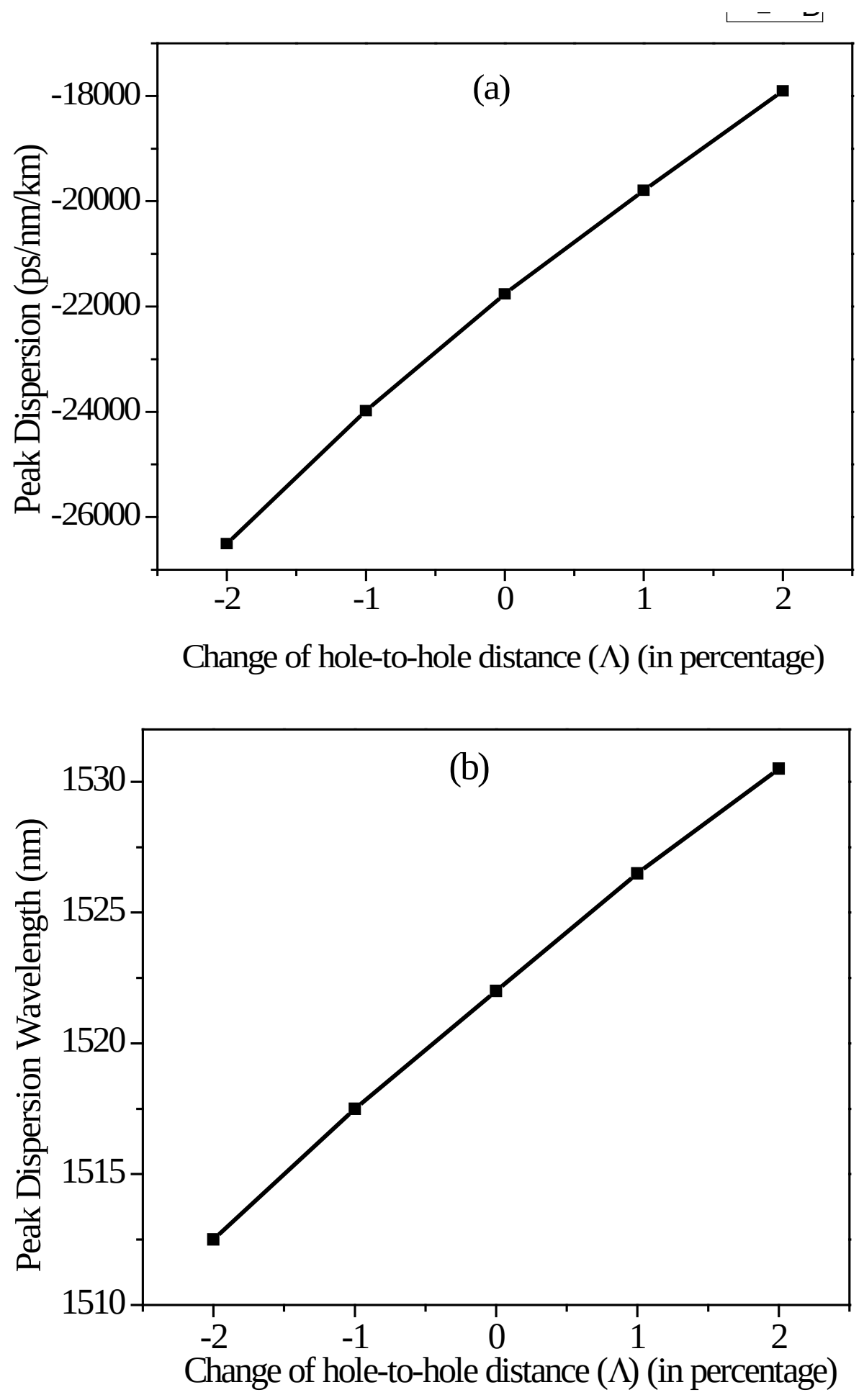

**Fig.** 8: variation of peak dispersion (a) and peak dispersion wavelength (b) for percentage change of hole-to-hole distance (Λ).

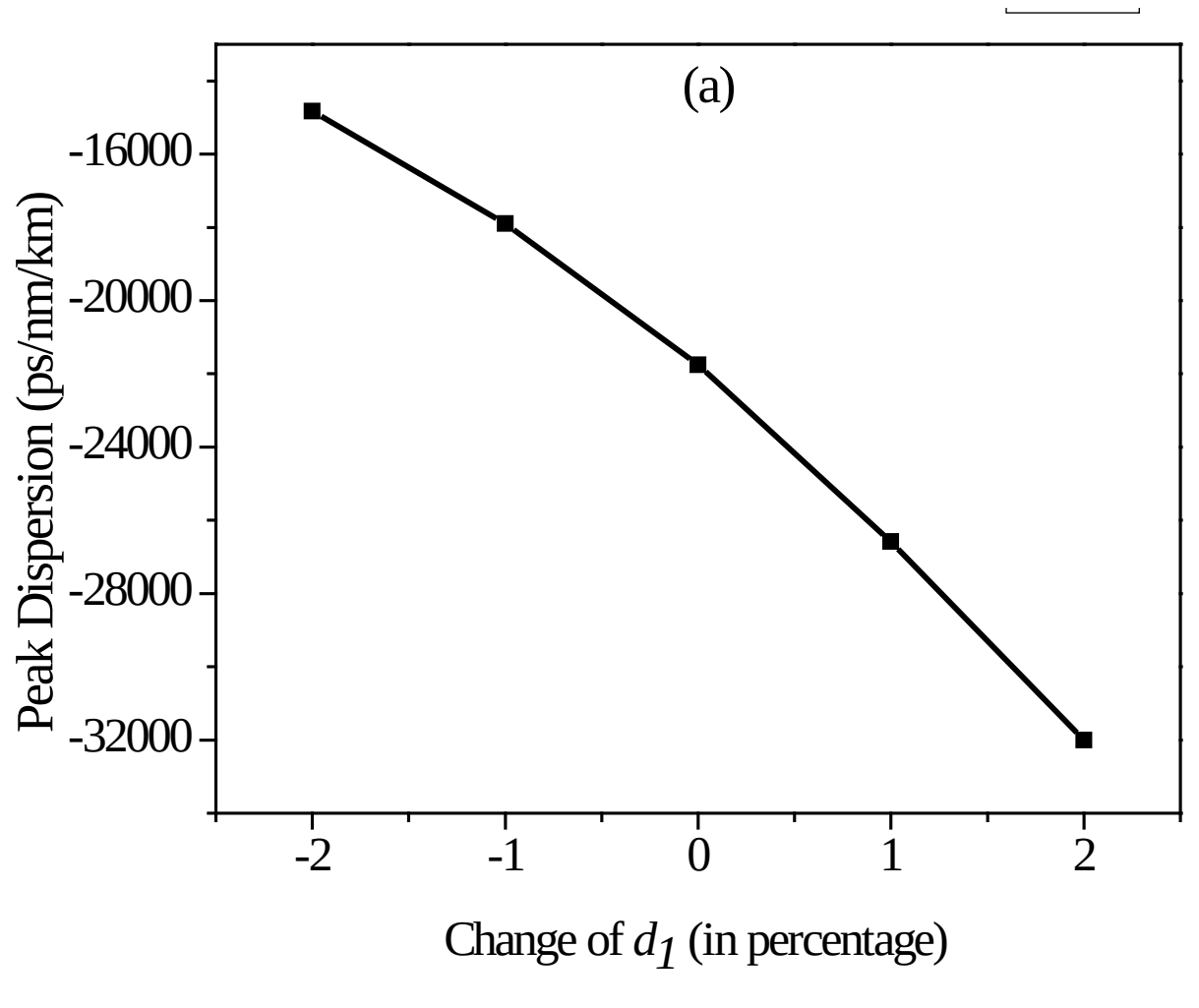

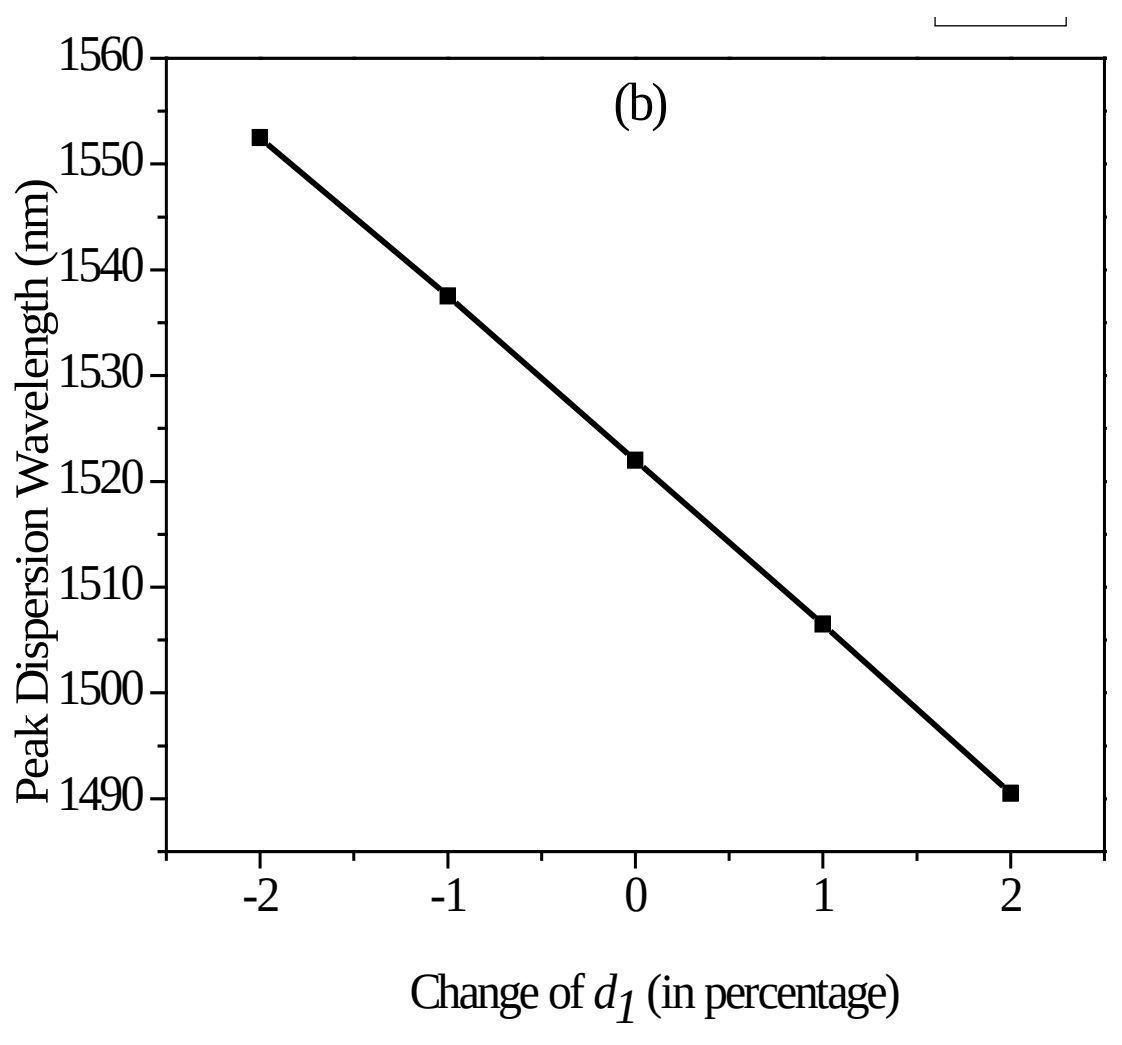

**Fig.** 9: Variation of peak dispersion (a) and peak dispersion wavelength (b) for percentage change of bigger air-hole diameter ($d_1$).

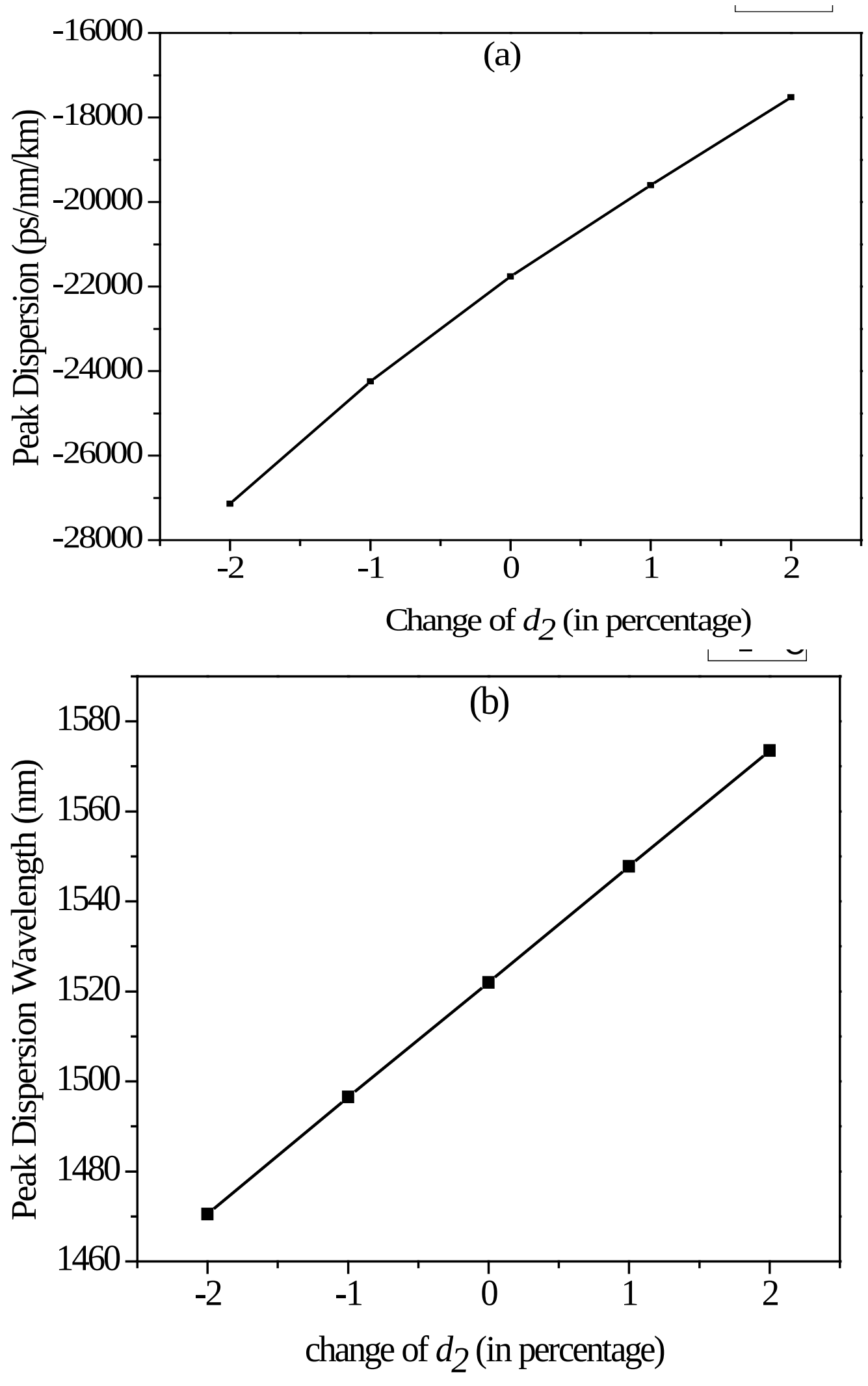

**Fig.** 10: Variation of peak dispersion (a) and peak dispersion wavelength (b) for percentage change of smaller air-hole diameter ($d_2$).

## 4. Ultra negative dispersion compensating dual core PCF

Based upon the above studies we changed the available parameters and we could achieve a very high negative dispersion of -47,500 *ps/nm/km* around 1550nm with Λ=1.4μm, $d_1$=1.196μm and $d_2$=0.59μm as shown in Fig. 11. This value of negative dispersion is the highest to the best of our knowledge as the value is more than twice than the earlier reported value with regular triangular lattice PCF [7-13] that can be drawn with regular stack and draw method. This optimized designed PCF will be very helpful in compensating the inline dispersion caused by SMF-28. As for example, the dispersion caused by existing SMF-28 is calculated through Eqn. (4) with a zero dispersion wavelength $\lambda_0$ of 1311.5 nm and a zero dispersion slope $S_0$ of 0.092 $ps\ nm^{-2}\ km^{-1}$ and is given by [**24**].

$$D(\lambda) = \frac{S_0}{4}[\lambda - \frac{\lambda_0^4}{\lambda^3}] \qquad (4)$$

At the wavelength of 1550nm the dispersion caused by the existing fiber is 17.378 *ps/nm/km*. So with a negative dispersion of -47,500*ps/nm/km* of the optimized design, we can compensate the dispersion caused by 27.3km of the existing inline optical fiber with only 10 meters of our optimized fiber. The confinement loss of the above PCF was calculated with Eqn. (3) and the value was found to be 0.089*dB/m* at the wavelength of 1550nm. To compensate the dispersion caused by 27.1km of SMF-28 as mentioned above, the total propagation losses will be 0.89*dB* for a propagation of 10 meter. So with this small length of the fiber not only the inline dispersion will be compensated but also the propagation losses caused by the fiber can also be reduced significantly. Similar requirements of short length fiber based devices can also be accomplished by present kind of devices. Effective area is another important parameter for long haul communications. The effective area variation for the optimized PCF has been presented in Fig. 12 for the entire C-band of wavelength. We have calculated the splice loss between the SMF-28 and the present fiber around 1550nm of wavelength and the value was found to be 2.2dB. The relatively higher loss can be attributed to the mismatch between the differences between the modal effective areas of the two fibers.

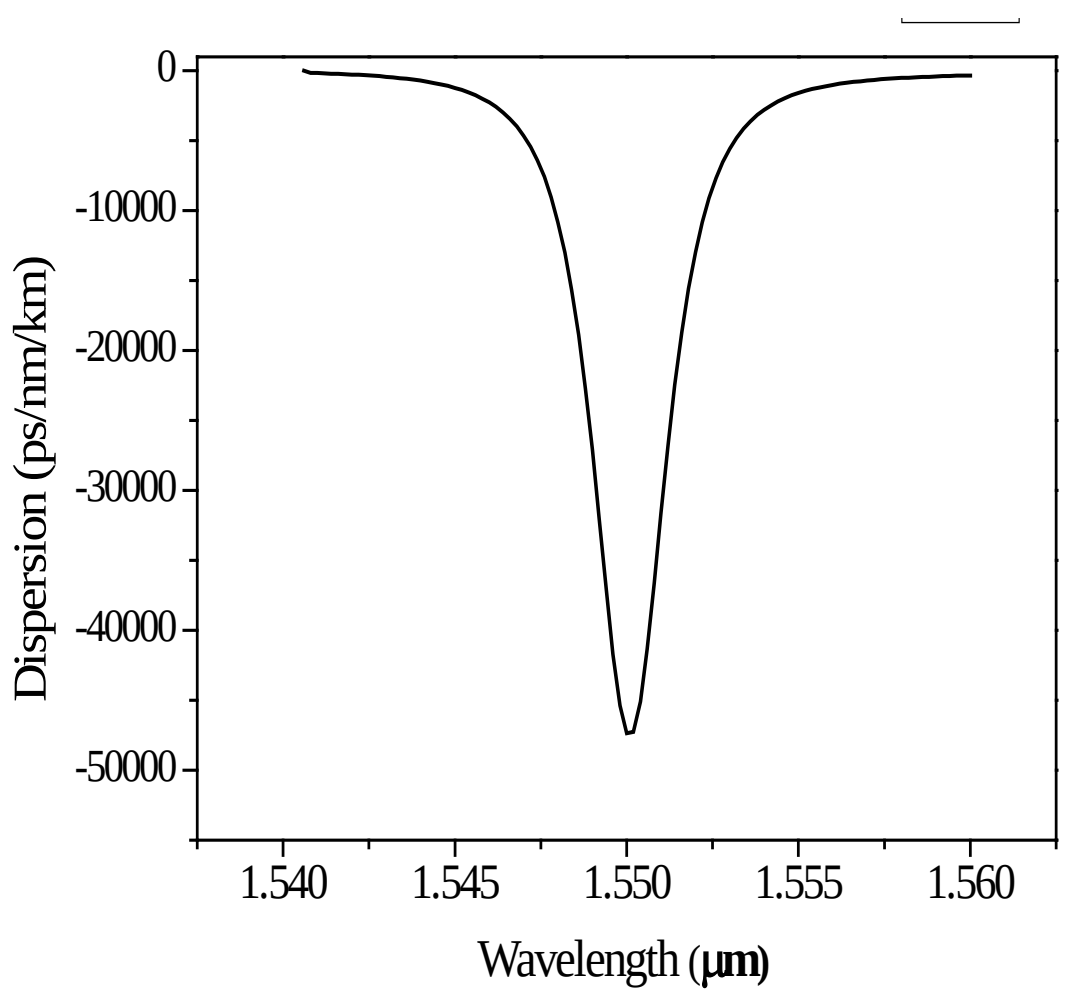

**Fig. 11:** Optimized dispersion value of -47,500*ps/nm/km* achieved with Λ=1.40μm with $d_1$=1.196μm and $d_2$ =0.59μm.

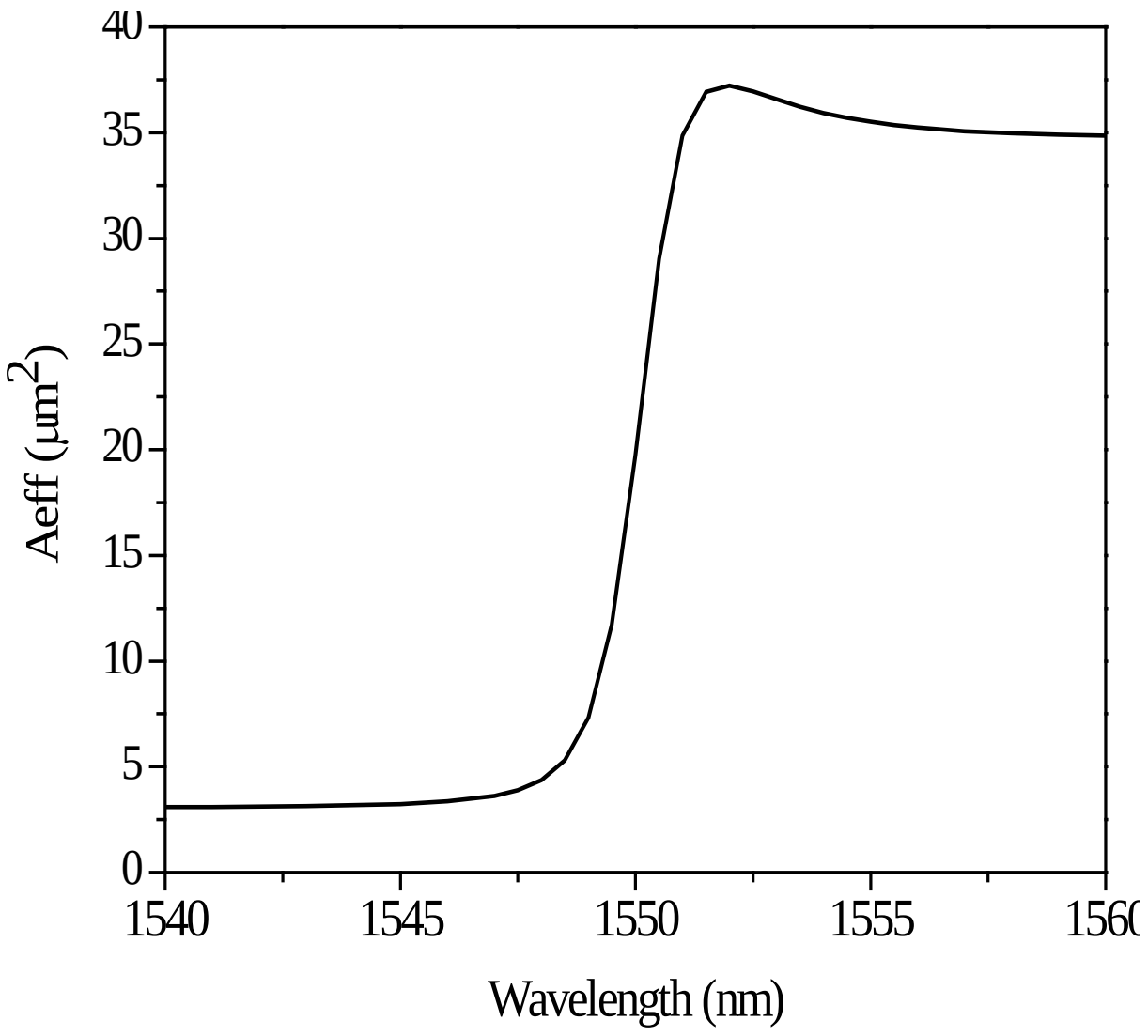

**Fig. 12**: Effective area variation of the optimized PCF.

## Conclusions:

In this paper we have theoretically investigated chromatic dispersion compensation property exhibited by square-lattice geometry of the PCFs based on pure silica. We have extensively studied the effect of different geometrical parameters upon dispersion towards achieving ultra-negative dispersion. We have shown that with an increase of bigger air-hole diameter, the peak dispersion is red-shifted with higher negative dispersion at the cost of narrower FWHM. While an increase of smaller air-hole diameter in the outer core again red-shifted the coupling wavelength but with smaller values of negative dispersion. Changing hole-to-hole distance has the effect of red-shifting the coupling wavelength with smaller values of negative dispersion. Based upon the above findings we could achieve an ultra negative dispersion of -47,000*ps/nm/km* around 1550nm of wavelength by properly changing the parameters. Our designed fiber will be very useful for dispersion compensation in long-haul data transmission some thousand times more than the available DCFs. The basic principle of power transform from inner core to the outer after the coupling can be applied for suppressing spontaneous emission after a particular wavelength.

## Acknowledgement:

The authors would like to thank Dr. Boris Kuhlmey, University of Sydney, Australia for providing valuable suggestions in understanding the software for designing and studying the properties of different structures. The authors acknowledge sincerely the Defence Research and Development Organization, Govt. of India and CRF of IIT Kharagpur for the financial support to carry out this research.

## References:

[1] J. Broeng, D. Mogilevstev, S. E. Barkou and A. Bjakle, "Photonic Crystal Fibers: a new class of optical waveguides" *Opt. Fiber Tech*. **5**, 305 (1999).

[2] J. C. Knight, "Photonic Crystal fibers," *Nature*, **424**, 8471(2003).

[3] L. P. Shen, W.-P. Huang, G. X. Chen, and S. S. Jian, "Design and optimization of photonic crystal fibers for broad-band dispersion compensation," *IEEE Photon. Tech. Lett.* **15,** 540 (2003).

[4] A. Huttunen and P. Törmä, "Optimization of dual-core and microstructure fiber geometries for dispersion compensation and large mode area," *Opt. Express* **13**, 627 (2005).

[5] G. Prabhakar, A. Peer, V. Rastogi, and A. Kumar," Large-effective-area dispersion-compensating fiber design based on dual-core microstructure," Appl. *Opt*. **52**, 4505 (2013).

[6] G. Ouyang, Y. Xu, and A. Yariv, "Theoretical study on dispersion compensation in air-core Bragg fibers," *Opt. Express* **10**, 899 (2002).

[7] T. D. Engeness, M. Ibanescu, S. G. Johnson, O. Weisberg, M. Skorobogatiy, S. Jacobs, and Y. Fink, "Dispersion tailoring and compensation by modal interactions in Omni Guide fibers," *Opt. Express* **11**, 1175 (2002).

[8] F. Poli, A. Cucinotta, M. Fuochi, S. Selleri, and L. Vincetti, "Characterization of microstructured optical fibers for wideband dispersion compensation," *J. Opt. Soc. Am. A* **20,** 1958 (2003).

[9] L. P. Shen, W.-P. Huang, and S. S. Jian, "Design of photonic crystal fibers for dispersion-related applications," *J. Lightwave Technol.* **21,** 1644 (2003).

[10] B. Zsigri, J. Laegsgaard, and A. Bjarklev, "A novel photonic crystal fibre design for dispersion compensation," *J. Opt. A: Pure Appl. Opt*. **6,** 717 (2004).

[11] Y. Ni, L. An, J. Peng, and C. Fan, "Dual-core photonic crystal fiber for dispersion compensation," *IEEE Photon. Technol. Lett.* **16,** 1516 (2004).

[12] F. Gérôme, J.-L. Auguste, and J.-M. Blondy, "Design of dispersion-compensating fibers based on a dual concentric-core photonic crystal fiber," *Opt. Lett*. **29**, 2725 (2004).

[13] S. Yang, Y. Zhang, X. Peng, Y. Lu, A. Xie, J. Li, W. Chen, Z. Jiang, J. Peng, and H. Li, "Theoretical study and experimental fabrication of high negative dispersion photonic crystal fiber with large area mode field,"*Opt. Express* **14**, 3015 (2006).

[14] X. Zhao, G. Zhou, S. Li, Z. Liu, D. Wei, Z. Hou, and L. Hou, "Photonic crystal fiber for dispersion compensation," *Appl. Opt*.**47**, 5190 (2008).

[15] S. Kim, C. S. Kee, D. K. Ko, J. Lee, and K. Oh, "A dual-concentric-core photonic crystal fiber for broadband dispersion compensation," *J. Korean Phys. Soc*. **49**, 1434 (2006).

[16] F. Poli, M. Foroni, M Bottacini, M. Fuochi, N. Burani, L. Rosa. A. Cucinotta, and S. Selleri, "Single mode regime of square-lattice photonic crystal fibers,"*J. Opt. Soc. A.* **22**, 1655(2005).

[17] A. H. Bouk, A. Cucinotta, F. Poli and S. Selleri, "Dispersion properties of square-lattice photonic crystal fibers," *Opt. Express* **12**, 941(2004).

[18] P. St. J. Russel, E. martin, A. Diez, S. Guenneau and A.B. Movchan, "Sonic band gaps in PCF performs: enhancing the iteration of sound and light," *Opt. Express* **11**, 2555(2003).

[19] M. K. Lee, P. S. Ma, I. K. Lee, H. W. Kim, and Y. Y. Kim, "Negative refraction experiments with guided shear-horizontal waves in thin phononic crystal plates" *App. Phy. Lett.*, **98**, 011909 (2011).

[20] G. P. Agrawal, Nonlinear Fiber Optics, 4th ed., Optics and Photonics Series (Academic, San Diego, Calif., 2007).

[21] CUDOS MOF utilities available online: http://www.physics.usyd.edu.au/cudos/mofsoftware/

[22] T. P. White, B. T. Kuhlmey, R. C. PcPhedran, D. Maystre, G. Renversez, C. M de Sterke and L. C. Botten, "Multipole method for microstructured optical fibers. I. Formulation" *J. Opt. Soc. Am. B*. **19**, 2322 (2002).

[23] B. T. Kuhlmey, T. P. White, R. C. PcPhedran, D. Maystre, G. Renversez, C. M de Sterke and L. C. Botten, "Multipole method for microstructured optical fibers. II. Implementataion and results." *J. Opt. Soc. Am. B.* **19**, 2331 (2002).

[24] ''Corning® SMF-28 CPC6 Single-Mode Optical Fibre,'' Product Information (Corning, Ithaca, N.Y., 1998).